# Do Smarter People Have Better Passwords? Yes, But…

**JV ROIG**, Advanced Research Center – Asia Pacific College

The National Institute of Standards and Technology (NIST) released new guidelines in June of 2017 that recommended new standards for managing and accepting user passwords. Among the new guidelines is a requirement that verifiers should check if a user's supplied password is compromised – that is, already listed in previous breach corpuses. Using a corpus of 320M breached passwords, the researcher collected information regarding Asia Pacific College students using breached passwords. Correlating these with academic performance data from each student's grade history, the researcher found that the students in the highest GPA tier had the lowest % of terrible passwords. The difference is not that large, however, which suggests that weak passwords aren't mainly because of any level of intelligence, nor should it be assumed that highly-intelligent users will have good passwords.

CCS Concepts • **General and reference** → Empirical studies; • **Security and privacy** → Security services → Authentication;

**KEYWORDS**
Password authentication; breached passwords; intelligence and password hygiene

## 1 INTRODUCTION

The National Institute of Standards and Technology (NIST) released new guidelines in June of 2017 that recommended new standards for managing and accepting user passwords [6]. Among the new guidelines is a requirement that verifiers should check if a user's supplied password is compromised – that is, already listed in previous breach corpuses [7]. These breach corpuses are huge lists of passwords that have been retrieved (mostly unethically and illegally) by black hats who exploited security holes in public-facing servers. The main problem here, of course, is getting a list of breached passwords.

Fortunately, Troy Hunt, an Australian security researcher, collected lists of breached data that have been published publicly, and then created an online service, "Have I Been Pwned?", where users can check if their usernames or passwords (or both) have been included in any past breach. After the NIST guidelines were published, he also publicly offered, for free download, a list of 320M passwords retrieved from past breaches he has recorded [9]. His motivation was to explicitly allow researchers and IT departments to implement the new NIST guideline of checking user passwords against these breached passwords, to be able to warn users if their password is unsafe (i.e., if it is found to match any password in the breached corpus).

Asia Pacific College (APC), where the researcher is currently employed, embarked on the implementation of this new guideline using the list from Troy Hunt, and along the way collected data about how many faculty, staff and students used passwords that were found in the list of 320M breached passwords from Troy Hunt.

The researcher, looking at the data collected from simply flagging whether a user had a breached password or not, had an idea: maybe we can use this data set to see if intelligence/smartness affects the use of terrible passwords? If we use the data solely of college students (disregarding staff and faculty), then we would already have a reasonable metric to categorize the intelligence level of each user: GPA (grade point average). This makes simple what would otherwise be the almost impractically complex problem of categorizing every user's intelligence level.

## 2 BACKGROUND

Passwords remain the most ubiquitous form of user authentication. Users are generally known to be terrible at managing passwords – terrible at creating them (prone to weak, insecure passwords) and terrible at remembering them. Convenience and human memory limitations make creation of strong passwords an unnatural tendenc, and left on their own with little guidance, users will inevitably revert to insecure password habits [1, 3]. Users aren't appreciative of overly complex and strict password policies [8], and these complex rules often have the side effect of encouraging coping behaviors from users that are not ideal, such as writing passwords down [4]. The seeming lack of care can be attributed to factors such as believing that the risks associated with their poor password hygiene are overblown or that they aren't personally at risk [5]. A Microsoft Research paper also found that users often reuse passwords (another negative password habit), with passwords reused for an average of 5.67 sites [2]. This same research shared that users were found to type passwords at an average of 8 instances per day.

These are all well-known difficulties between users and the password authentication mechanism that they should be more careful of. New guidelines, such as the previously mentioned NIST Digital Identity Guidelines of June 2017, have taken into account behavioral factors into refining password policies.

How about intelligence? Does intelligence level affect password hygiene, and perhaps be a factor that should be taken into account by password policies?

## 3 EXPERIMENT DESIGN AND METHODOLOGY

### 3.1 Collecting Data Regarding Relative Password Quality Per Intelligence Level

This experiment benefits from an existing research infrastructure within APC that is continually collecting data. This existing research infrastructure, which we shall call "PWR" (Password Research) for ease of reference, provided the researcher with a list of users within APC, plus a marker for each user, labeled "Compromised", which contains a value that is either "TRUE" or "FALSE": "TRUE" means the user's password was found to exist in Troy Hunt's list of 320M passwords that are unsafe and therefore to be considered as compromised (passwords retrieved from past exploits and are known by malicious actors), "FALSE" means the password was not found. The design of PWR is detailed in the next section.

With a list of usernames and the compromised password marker, the researcher filtered out all users that were not college students – usernames of Employees and Senior High students were removed. With assistance from APC's IT team, the researcher created an automated script that hooked into APC's school informaiton system to create a text file that matches each username with their corresponding GPA. To maintain anonymity and privacy, the usernames in this list are not the actual usernames of the students; rather, they are hashes of the actual username, so that the researcher and the IT team does not actually know the identities of the students. The automated script used a lookup table (provided by the PWR infrastructure) in order to be able to match the GPA to the hashed usernames.

At the end of the process above, the researcher then has a list that contains the hashed usernames (anonymized users, but all known to be college students), the compromised password marker of TRUE/FALSE, and the GPA. From here, the researcher processed the text file in buckets of GPA ranges to see how the smarter students compared to their lower GPA brethren in terms of usage of very unsafe passwords.

### 3.2 The PWR Infrastructure Design

The PWR infrastructure is composed of three main components. Together, these components provide APC researchers with insight into the password hygiene of APC users, with the data derived here used to assess and refine the organization's password policies and password education effort.

The first component of PWR is Troy Hunt's list of 320M hashed passwords. The list is entirely in SHA1 hashes, no plaintext passwords to avoid being weaponized [https://www.troyhunt.com/introducing-306-million-freely-downloadable-pwned-passwords/]. As of the time the PWR infrastructure was created, Troy Hunt's release consisted of 3 separate files – the original release plus two updates. Combined, the list of hashed passwords is well over 10GB in size. Searching through for a hash in such a large file is slow. In order to speed up searches for a hash, the author split the hash list into 256 different files, with each hash going to a specific file based on the first two characters. Hashes that start with "aa", for example, were sent to the "aa.txt" file. Since the hashes are rendered as a string of hexadecimal characters, the files start from "00.txt", with the last file being "ff.txt". Searching for a particular hash means determining first what file to search in based on the first two characters of the hash, and then running the Linux utility *grep* on that file to search for the specific hash. This delivers a speedup of two orders of magnitude, which will be needed for the PWR use case as described further below.

The second component is a set of scripts that hook into APC's existing school information system (APCIS), which is PHP-based. The scripts are also in PHP to make them immediately compatible and easy to integrate. These PWR scripts add hooks to the login page of APCIS, so that users who log in will have their passwords checked if they are compromised or not based on the first component described above. The passwords are first hashed using SHA1 (to make them compatible with the Troy Hunt password list), and then that hash is checked against the list of hashes provided by component one. If the hash is found in the list, that means the user's password is part of the breach corpus, and it is marked as compromised. The second component also provides a small library of tools that anonymize the information. Of note is a special hash function that is used so that usernames are transformed into a hash, so that users become unidentifiable.

The third component of PWR is an SQLite3 database that stores the information that the PWR component two produces. To maintain anonymity of the data stored, PWR does not log the actual username. As described above, PWR transforms the username into a hash, and this hashed username is what is stored into the PWR database. Other relevant information, such as user type ("Employee", "College Student", "Senior High Student"), age bracket, gender, course / department, and password length are also stored along with the hashed username and the compromised password marker.

Overall, PWR was designed to be compliant with NIST's latest guideline of checking a user's password against breach corpuses. The data collected by PWR is used by APC's research team for refinement of password policies and strengthening the community's password hygiene.

## 4 ANALYSIS AND RESULTS

The PWR data consisted of 1,252 APC users, all of which are college students (non-college students were already filtered out by the author). This represents 91.52% of all active college students. PWR data also show that 215 of these users (17.17% of the total) were marked as "compromised", meaning these 215 use passwords that are in the PWR breach corpus.

For reference, APC's grading system uses 4.0 as the highest, with 1.0 as the minimum passing grade.

Table 1: Statistics of users with unsafe passwords categorized per minimum GPA.

| Min GPA | # of Students | Compromised Passwords | % of Compromised Passwords | Remarks |
| --- | --- | --- | --- | --- |
| 3.5 | 39 | 5 | 12.82% | ~13% of students with a GPA of 3.5 and above use unsafe passwords |
| 3.0 | 242 | 37 | 15.29% | 3.0 and up (including all the 3.5 students from above) make the % jump closer to the population average. |
| 2.5 | 688 | 103 | 14.97% | The % here goes slightly down. This means the 2.5-3.0 GPA students contribute a lower % compromised compared to the 3.0-3.5 GPA students. |
| 2.0 | 1152 | 195 | 16.93% | 17% of students with a GPA of 2.0 to 4.0 use unsafe passwords. Not surprising that we are approaching the population average now that we're counting majority of the population already |
| 1.5 | 1251 | 214 | 17.11% | |
| 1.0 | 1252 | 215 | 17.17% | The single student who had a lower than 1.5 GPA also happened to use an unsafe password |

Table 1 shows an analysis of the distribution of compromised passwords according to minimum GPA. This is not per bucket / per tier. The last row, for example, contains the entire population (1,252) since it is the lowest grade – therefore, everybody qualifies in this criteria. What this shows how slicing the population into a certain GPA range compares to the population average of 17.17% compromised passwords. This shows us, for example, that if we only take into account students with a GPA of at least 3.5, only 12.82% of them use compromised passwords, which compares favorably to the population average of 17.17%. Looking at students with a minimum GPA of 3.0 results in 15.29% compromised passwords, which is significantly closer to the population average.

Table 2: A per GPA-tier view of percentage of compromised passwords.

| GPA Tier | # of Students | Compromised Passwords | % of Compromised Passwords | Remarks |
| --- | --- | --- | --- | --- |
| 3.5 - 4.0 | 39 | 5 | 12.82% | Students with a GPA between 3.5 and 4.0 had the lowest % of unsafe passwords |
| 3.0 – 3.49 | 203 | 32 | 15.76% | Honor students – between 3.0 to 3-5 – fared slightly worse than the tier below |
| 2.5 – 2.99 | 446 | 66 | 14.80% | |
| 2.0 – 2.49 | 464 | 92 | 19.83% | These students had the worst number of unsafe passwords in terms of percentage |
| 1.5 – 1.99 | 99 | 19 | 19.19% | Second-worst tier in terms of percentage |
| 1.0 – 1.49 | 1 | 1 | 100.00% | The single student who had a lower than 1.5 GPA also happened to use an unsafe password |

Table 2 shows the same data in a slightly different view – instead of aggregating students together gradually through minimum GPA, this time they are separated into 6 GPA tiers (from 1.0 to 4.0 in 0.5 increments). As Table 1 seemed to indicate, Table 2 makes it clear that the 2.5-2.99 GPA tier has a slightly better percentage compared to the higher tier of 3.0-3.49.

Whether viewed as grouped per minimum GPA as in Table 1, or as GPA tiers in Table 2, it is obvious that the more intelligent students (at least, if we take for granted for now that a higher GPA means more intelligent) do outperform their lower-GPA brethren. Even discounting the lowest tier since it only has a sample size of 1, the final two tiers do have significantly higher percentages of compromised password use compared to the higher tiers, and outperform the population average of 17.17%

The PWR dataset used here also indicates that the mean APC password length is 11.2 characters with a mode of 10 characters. Over 98% of users have 8 characters or more, over 50% have 10 or more characters, and over 25% use at least 12 characters. This means the usage of short passwords is an almost non-existent problem for APC users. With the prevalence of long passwords (8+ characters, as per NIST recommendation), a password that is not found in the PWR hash lists has a good chance of being a strong password.

## 5   CONCLUSION AND RECOMMENDATIONS

This analysis seems to show that smarter people do seem to have better password hygiene. While quantitative results such as these are always helpful as insights for refining password policies, it is important to note several caveats in this study:

> 1.) *This assumes that higher GPA means smarter.* While this may generally be the case, this is far from a foregone conclusion. Smartness or intelligence is a complex subject, and the measurement of intelligence is not something that is trivial and universally accepted. A different study that has access to other measures of intelligence – such as standardized aptitude tests – to combine with GPA may yield further insightful results
>
> 2.) *This only classifies password as good or bad based on whether it is in the list of known bad passwords.* Being found in the hash list of the PWR infrastructure definitely means the password is unsafe. However, the reverse is not automatically true – just because the password is not list of known bad passwords doesn't mean it is a perfectly safe one. While this is likely to be true here due to the prevalence of long passwords among APC users, this isn't a guarantee. A different study that has access to more password parameters for classifying the strength of passwords may yield further insightful results. (The author would note here that within the current PWR infrastructure, it is impossible to make a password strength analysis based on the character composition of the password. The PWR infrastructure does not handle or store plaintext passwords)
>
> 3.) *The sample population here is limited and may be biased.* The experiment is limited to APC college students. There may be a certain bias inherent in the population, so repeating the experiment in a different schools may yield further insightful results.

Do smarter people have better passwords? It seems so, at least among college students, and assuming GPA is a good measure of intelligence, and that the APC college population is a good representation of users in general. This shouldn't be taken as the end-all or be-all of whether smarter people have better passwords, but merely one interesting data point in what could be an interesting series of further experiments.